\documentclass[journal]{IEEEtran}
\usepackage{graphicx}
\usepackage{amsmath}
\usepackage{amsfonts}
\usepackage{algorithm}
\usepackage{algorithmic}
\usepackage{makecell}
\usepackage{booktabs}
\setcellgapes{3pt}

\setcellgapes{3pt}

\usepackage[justification=centering]{caption}
\usepackage{color}

\usepackage{cite}

\ifCLASSINFOpdf
\else
\fi
\begin{document}
%
\title{ Deep Learning Based Joint Resource Scheduling Algorithms for Hybrid MEC Networks}

\author{Feibo Jiang, Kezhi Wang, Li Dong, Cunhua Pan, Wei Xu and Kun Yang
	\thanks{
		This work was supported in part by the National Natural Science Foundation
		of China under Grant no. 41604117, 41904127, 41874148, 61620106011, 61572389 and 61871109. This work was also supported in part by the Royal Academy of Engineering under the Distinguished Visiting Fellowship scheme (DVFS21819$\backslash$9$\backslash$7) and by Scientific Research Fund
		of Hunan Provincial Education Department in China under Grant no. 18A031}
		\thanks{Feibo Jiang (jiangfb@hunnu.edu.cn) is with Hunan Provincial Key Laboratory of Intelligent Computing and Language Information Processing, Hunan Normal University, Changsha, China, Kezhi Wang (kezhi.wang@northumbria.ac.uk) is with the department of Computer and Information Sciences, Northumbria University, UK, Li Dong (Dlj2017@hunnu.edu.cn) is with Key Laboratory of Hunan Province for New Retail Virtual Reality Technology, Hunan University of Technology and Business, Changsha, China, Cunhua Pan (Email: c.pan@qmul.ac.uk) is with School
		of Electronic Engineering and Computer Science, Queen Mary University
		of London, London, E1 4NS, UK, Wei Xu (wxu@seu.edu.cn) is with NCRL, Southeast University,
		Nanjing, China,	
		Kun Yang (kunyang@essex.ac.uk) is with the School of Computer Technology and Engineering, Changchun Institute of Technology, Changchun, China and also with the School of Computer Sciences and Electrical Engineering, University of Essex, CO4 3SQ, Colchester, UK. }
			\thanks{Corresponding authors: Kezhi Wang and Kun Yang }
}

\markboth{Submitted for Review}%
{Shell \MakeLowercase{\textit{et al.}}: Bare Demo of IEEEtran.cls for IEEE Journals}
%



\maketitle

\begin{abstract}

In this paper, we consider a hybrid mobile edge
computing (H-MEC) platform, which includes ground stations (GSs), ground vehicles (GVs) and unmanned aerial vehicle (UAVs), all with mobile edge cloud installed to enable user equipments (UEs) or Internet of thing (IoT) devices with intensive computing tasks to offload. Our objective is to obtain an online offloading algorithm to minimize the energy consumption of all the UEs, by jointly optimizing the positions of GVs and UAVs, user association and resource allocation in real-time, while considering the dynamic environment. To this end, we propose a hybrid deep learning based online offloading (H2O) framework where a large-scale path-loss fuzzy c-means (LSFCM) algorithm is first proposed and used to predict the optimal positions of GVs and UAVs. Secondly, a fuzzy membership matrix U-based particle swarm optimization (U-PSO) algorithm is applied to solve the mixed integer nonlinear programming (MINLP) problems and generate the sample datasets for the deep neural network (DNN) where the fuzzy membership matrix can capture the small-scale fading effects and the information of mutual interference. Thirdly, a DNN with the scheduling layer is introduced to provide user association and computing resource allocation under the practical latency requirement of the tasks and limited available computing resource of H-MEC. In addition, different from traditional DNN predictor, we only input one UE’s information to the DNN at one time, which will be suitable for the scenarios where the number of UE is varying and avoid the curse of dimensionality in DNN.

\end{abstract}


%
\IEEEpeerreviewmaketitle

\section{Introduction}

%
%
%
%
Computation intensive applications, such as face recognition, language processing, online gaming, augmented reality (AR) and virtual reality (VR), have been fast developing and increasingly outgrowing the limited capabilities of devices\cite{6249269}. Thanks to the recent advancement of mobile edge computing (MEC) technology, the gap between the limited amount of resource in devices and the demands for better experience is being reduced\cite{6616113,8063331}. With the help of MEC, the UE can offload the intensive computations to the nearby edge servers to save energy consumption and increase the computational capacity \cite{8353131,8713801}. However, different from the traditional cloud in data center, edge cloud may be implemented by the router, switches, which may have some free computing resource and are closer to the users. Recently, vehicle and UAV \cite{8647789} based MEC has also been proposed. Due to limited amount of the computing resource in MEC and finite physical bandwidth of wireless channels, task admission control and resource allocation are normally required, especially in the presence of a large number of delay-sensitive tasks. The problem is generally formulated as mixed integer nonlinear programming (MINLP). To tackle the MINLP problem, branch-and-bound algorithms\cite{1674939} and dynamic programming \cite{bertsekas1995dynamic} are normally used to obtain the globally optimal offloading solution. However, the search spaces of both methods increase exponentially with the network size and are computationally prohibitive for large-scale MEC networks. To reduce the computational complexity, heuristic local searching methods are proposed. For instance, \cite{8334188} proposed a coordinate descent (CD) method that searches along one binary variable at a time. A similar heuristic search method for multi-server MEC networks was studied in \cite{8533343}, which iteratively adjusts binary offloading decisions. Another widely adopted heuristic method is through convex relaxation, e.g., by relaxing integer variables to be continuous between 0 and 1 \cite{7524497} or by approximating the binary constraints with quadratic constraints \cite{7914660}.

Nonetheless, on one hand, the solution quality of reduced-complexity heuristics is not guaranteed. On the other hand, both searching-based and convex relaxation methods often require a large amount of iterations for an algorithm to reach a satisfying local optimum. Hence, they are unsuitable for real-time processing in fast changing environment, as the optimization problem needs to be re-solved once the number and position of UEs have varied significantly.

Recently, some artificial intelligence methods have emerged as effective tools for enhancing MECs. In \cite{8490683}, the energy-efficient computation offloading management scheme in the MEC system with small cell networks (SCNs) is proposed, and a hierarchical genetic algorithm (GA) and particle swarm optimization (PSO)-based heuristic algorithm are designed to solve this problem. In \cite{jiang2018predicted}, a deep learning (DL) algorithm based on multi-long and short-term memory (LSTM) networks is proposed to forecast the traffic of small base stations (SBSs), on the basis of which an offline mobile data offloading strategy obtained through on cross-entropy is presented. In \cite{7875131}, a conceptor-based echo state network is proposed to predict content request distribution of users and its mobility pattern when the network is available. Based on the prediction results, the optimal positions of UAVs and the content to cache at UAVs can be obtained. In \cite{Huang2018}, a distributed deep learning offloading algorithm is introduced in MEC networks, where multiple parallel DNNs are trained separately and applied to make offloading decisions cooperatively. In \cite{8664596}, an emerging deep neural network technique is used in the mobile crowd sensing (MCS). The proposed technique employs convolutional neural network for feature extraction, and then directs the sensing and movement of UEs under the guidance of the distributed multi-agent deep deterministic policy gradient method. In \cite{8270639}, a deployment strategy for the distributed multi-layer convolutional neural network is presented. The strategy divides the convolutional neural network into two parts: the preprocessing part and the classification part. The preprocessing part is deployed on the edge server for feature extraction and compression of the image data so as to reduce the data transmission between the edge server and the cloud server.

Nonetheless, on one hand, the heuristic computation algorithms have excellent global search ability and high calculation accuracy, but they need long computing time. On the other hand, the supervised learning algorithms have outstanding prediction and reasoning capabilities, but they require a large amount of labelled training data.

In this paper, we consider a hybrid mobile edge computing (H-MEC) platform, where there are ground stations (GSs), ground vehicles (GVs) and unmanned aerial vehicle (UAVs), all with edge cloud enhanced, which can enable UEs with computational intensive tasks to offload. We aim to obtain an online offloading algorithm to minimize the energy consumption of all the UEs, by jointly optimizing the positions of GVs and UAVs, user association and resource allocation in real time, while considering the dynamic environment. Towards this end, we propose a hybrid deep learning based online offloading (H2O) framework to achieve the above targets. Compared with the existing integer programming and learning based methods, we have the following novel contributions:
\begin{itemize}

\item We first introduce a large-scale path-loss fuzzy c-means (LS-FCM) clustering algorithm to locate the positions of UAVs and GVs, which has two improvements compared to the traditional FCM: First, it can fix some cluster centers denoted as GS positions and not allow them to participate in the iteration process. Second, it introduces the  large-scale path-loss component to replace distance in clustering process.

\item We then introduce a fuzzy membership matrix U-based particle swarm optimization (U-PSO) algorithm, which can solve the task admission and resource allocation problem with different initial states. This procedure is repeated until enough samples are collected. PSO can solve the complex MINLP problems precisely and provide high quality labeled samples to the DNN for offline training. Additionally, a U-based roulette wheel selection strategy is applied to guide the initial stage of PSO and provide high quality initial solution for accelerating convergence of PSO, where the fuzzy membership matrix can capture the small-scale fading and the information of mutual interference.

\item The DNN is applied for real-time decision-making. The training stage is done by the collected samples from U-PSO, by applying the continuous membership information as the input for offloading action generation and resource allocation. This hybrid mechanism keeps the advantages of the PSO in finding global optimal solutions, while speeding up the decision-making through DNN. Besides, to generate an offloading action, the proposed H2O framework only needs to input the membership information of one UE each time. Compared to some conventional deep learning methods that require to input the information of all UEs, H2O is computationally feasible and efficient in large-size networks with different number of UEs, which is suitable for continually dynamic scenarios. Moreover, the online new input and output of the DNN will be collected, recalculated and stored back to the sample memory database, which is very important for improving the performance of DNN in real scenarios.

\item We further develop an additional scheduling layer of DNN to check if the constraints are guaranteed. Also, admission control is conducted in this layer. The H2O framework is suitable for solving large-scale MINLP problem in real-time. We finally evaluate the proposed H2O framework under extensive numerical studies. The experiment results of the proposed H2O are compared with several different kinds of benchmark solutions, which demonstrates the feasibility and effectiveness of our framework. The simulation results have also shown that our solutions have better computational efficiency and accuracy. This can make real-time and optimal design feasible in the H-MEC networks even in a fast changing environment.

\end{itemize}

The remainder of this paper is organized as follows. In Section II, we describe the system model and problem formulation. We introduce the detailed designs of the H2O algorithm in Section III. Numerical results are presented in Section IV. Finally, the paper is concluded in Section V.




\section{System model and problem formulation}
\begin{figure}[htpb]
  \centering
	\includegraphics[width=8.8cm]{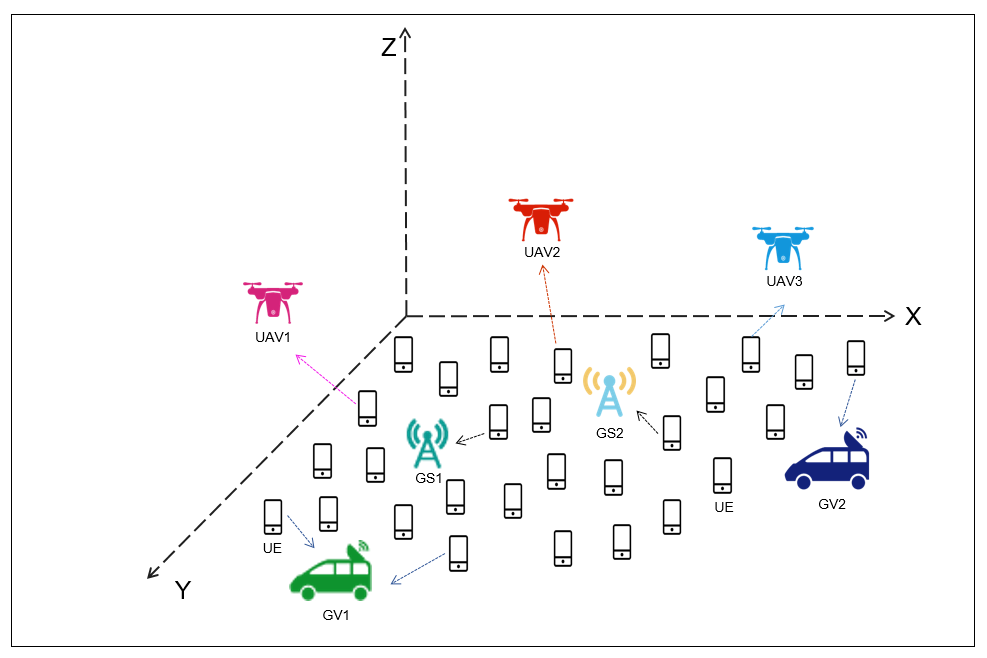}
	\caption{The H-MEC network.}
  \label{fig:fig1}
\end{figure}

Fig. \ref{fig:fig1} shows our proposed H-MEC networks with several GSs, GVs and UAVs, all of which have edge server enhanced. The locations of GSs are assumed to be fixed whereas the locations of GVs and UAVs can be optimized.

We consider that there are $N$ UEs randomly distributed in the ground, each of which has a computation task to be executed. The set of UEs is denoted as $\mathcal{N}=\{1,2,\cdots,N\}$. Consider there are $J$ UAVs, $K$ GVs and $M$ GSs, which can enable the UEs to offload the tasks. The sets of UAVs, GVs and GSs are denoted as $\mathcal{J}=\{1,2,\cdots,J\}$, $\mathcal{K}=\{1,2,\cdots,K\}$ and $\mathcal{M}=\{1,2,\cdots,M\}$, respectively.
We use variables $a_i^L$, $a_{i,j}^{MEC}$  to denote the possible places where the UEs can execute the tasks, where $a_i^L$ and $a_{i,j}^{MEC}$ denote the local executing, and offloading to the H-MECs (including UAVs, GVs and GSs), respectively. Then we have the following:
\begin{align}\label{eq:Shi1}
  a_{i}^L=\{0,1\},\forall i\in \mathcal{N} \notag\\
  a_{ij}^{MEC}=\{0,1\}, \forall i\in \mathcal{N},\forall j\in \{\mathcal{J},\mathcal{K},\mathcal{M}\}
\end{align}
where $a_i^L=1$ means that the $i$-th UE decides to execute the task itself, and $a_i^L=0$ otherwise, $a_{ij}^{MEC}=1$ means that the $i$-th UE decides to offload the task to the $j$-th UAV (when $j\in \mathcal{J}$), or to the $j$-th GV (when  $j\in \mathcal{K}$), or to the $j$-th GS (when $j\in \mathcal{M}$), and $a_{ij}^{MEC}=0$  otherwise. It is assumed that each task can be executed at most one place. Then, we have
\begin{equation}\label{eq:Shi2}
 a_{i}^L+\Sigma _{j\in \{\mathcal{J,K,M}\} }\ a_{ij}^{MEC}=1, \forall i\in \mathcal{N}.
\end{equation}

Similar to \cite{Wang2015Joint}, we assume that the $i$-th UE has the computational intensive task $Q_i$ to be executed as follows:
\begin{equation}\label{eq:Shi3}
Q_i=(F_i,D_i,T^{req}),\forall i\in \mathcal{N}
\end{equation}
where $F_i$ describes the total number of the CPU cycles of $Q_i$ to be computed, $D_i$ denotes the data size transmitting to the H-MECs if the offloading action is decided and $T^{req}$ is the latency constraint or quality of service (QoS) requirement of this task. Without loss of generality, in this paper, we consider that all the tasks have the same time requirement $T^{req}$. $D_i$ and $F_i$ can be obtained by using the approaches provided in \cite{6253581}. If the task is selected to offload, the execution time of the task is given by:
\begin{equation}\label{eq:Shi4}
T_{ij}^{C}=\frac{F_i}{f_{ij}}
\end{equation}
where $f_{ij}$ is the computation capacity of the $j$-th H-MEC providing to the $i$-th UE. Also, the time to offload the data is given by:
\begin{equation}\label{eq:Shi5}
T_{ij}^{Tr}=\frac{D_i}{r_{ij}}
\end{equation}
where $r_{ij}$ is the offloading data rate from the $i$-th UE to the $j$-th place. Then, we can have:
\begin{equation}\label{eq:Shi6}
T_{ij}^{Tr}+T_{ij}^{C}=\frac{D_i}{r_{ij}}+\frac{F_i}{f_{ij}} \le T^{req}
\end{equation}
which means that each task must meet the latency requirement. If this task is executed in UE itself, one has
\begin{equation}\label{eq:Shi7}
\frac{F_i}{f_{i}^{L}} \le T^{req}
\end{equation}
where $f_i^L$ is the local executing capacity. Therefore, one can have
\begin{equation}\label{eq:Shi8}
 a_i^{L}\frac{F_i}{f_{i}^{L}}+\sum _{j\in \{\mathcal{J,K,M}\} }\ a_{ij}^{MEC}(\frac{D_i}{r_{ij}}+\frac{F_i}{f_{ij}})\le T^{req},\forall i\in \mathcal{N}.
\end{equation}

Also, assume that the computing capacities in the UEs, UAVs, GVs and GSs are limited as
\begin{align}\label{eq:Shi9}
   a_{i}^Lf_{i}^{L}\le F_{i,max}^{L},\forall i\in \mathcal{N}\notag\\
\sum_{i\in \cal N }\ a_{ij}^{MEC}f_{ij}\le F_{j,max}^{MEC},\forall j\in \{\mathcal{J,K,M}\}
\end{align}
where $F_{i,max}^{L}$ is the local computational capability of the $i$-th UE, $F_{j,max}^{MEC}$ is the remote computational capability of the $j$-th H-MEC. The power consumption in the $i$-th UE can be given by
\begin{equation}\label{eq:Shi10}
P_{i}^{ue}=\begin{cases}
  \sum_{j\in \{\mathcal{J,K,M}\}}a_{ij}^{MEC} P_{ij}^T, &\,\text{if offloading} \\
  P_i^E,& \text{if local execution}
\end{cases}
\end{equation}
where $P_{ij}^T$ is the transmitting power from the $i$-th UE to the $j$-th H-MEC and $P_{i}^E$ is the execution power in the $i$-th UE if UE conducts the task itself and
\begin{equation}\label{eq:Shi11}
P_i^E=k_i(f_i^L)^{v_i},\forall i\in \mathcal{N}
\end{equation}
where $k_i\ge 0$ is the effective switched capacitance and  $v_i\ge 1$ is the positive constant. To match the realistic measurements, we set $k_i=10^{-27}$ and $v_i=3$\cite{Miettinen2010Energy}.

Assume that the coordinates of the $i$-th UE, the $j$-th UAV, the $j$-th GV and the $j$-th GS are $(x_i,y_i)$,  $(X_j^{UAV}, Y_j^{UAV}, H_j^{UAV})$,  $(X_j^{GV},Y_j^{GV})$, and $(X_j^{GS},Y_j^{GS})$, respectively. Then, the horizontal distance between the $i$-th UE and the $j$-th UAV is given by
\begin{equation}\label{eq:Shi12}
R_{ij}^{UAV}=\sqrt{(X_j^{UAV}-x_i)^2+(Y_j^{UAV}-y_i)^2},\forall i\in \mathcal{N},\forall j\in \mathcal{J}.
\end{equation}

If UEs decide to offload the task to UAVs, the data rate can be given as
\begin{equation}\label{eq:Shi13}
r_{ij}=B \text{log}_2\left(1+\frac{P_{ij}^Th_{ij}^{UAV}}{\sigma^2}\right),  \forall i\in \mathcal{N},\forall j\in \mathcal{J}
\end{equation}
where $B$ is the channel bandwidth, $h_{ij}^{UAV}$ is the channel quality information (CQI), denoted as $h_{ij}^{UAV}=\alpha_j^{UAV}\left(\sqrt{H_j^{UAV^2}+R_{ij}^{UAV^2}}\right)^{-2}$, $\alpha_j^{UAV}$ is the small-scale fading component, $\left(\sqrt{H_j^{UAV^2}+R_{ij}^{UAV^2}}\right)^{-2} $ is the large-scale path-loss component \cite{6787113}. Note that we assume that the users offload their tasks to UAV via orthogonal frequency division multiplexing (OFDM) channels, which means that there is no interference between each other.
Similar to \cite{8103781}, we assume that horizontal coverage of UAV is constrained by the following
\begin{equation}\label{eq:Shi14}
 R_{ij}^{UAV}\le H_j^{UAV}\text{tan}\phi_j^{UAV}, \forall i\in \mathcal N,\forall j\in \mathcal J
\end{equation}
where $\phi_j$ can be decided by the angle of antenna in the UAV \cite{8103781}.

Then, the horizontal distance between the $i$-th UE and the $j$-th GV is  \begin{equation}\label{eq:Shi15}
R_{ij}^{GV}=\sqrt{(X_j^{GV}-x_i)^2+(Y_j^{GV}-y_i)^2},\forall i\in \mathcal{N},\forall j\in \mathcal{K}.
\end{equation}

If UEs decide to offload to the GVs, the data rate can be given as:
\begin{equation}\label{eq:Shi16}
\begin{split}
r_{ij}=B\text{log}_2\left(1+\frac{P_{ij}^Th_{ij}^{GV}}{\sum_{k\in \mathcal{N} k\not=i}P_{kj}^Th_{kj}^{GV}+\sigma^2}\right),  \\ \forall i\in \mathcal{N},\forall j\in \mathcal{K}
\end{split}
\end{equation}
where we assume the UEs share the same channel when they decide to offload to the same GV and there is interference between them. $h_{ij}^{GV}=\alpha_j^{GV}\left(R_{ij}^{GV}\right)^{-2}$, $\alpha_j^{GV}$ is the small-scale fading component and $\left(R_{ij}^{GV}\right)^{-2}$ is the large-scale path-loss component.

The horizontal distance between the $i$-th UE and the $j$-th GS is as
\begin{equation}\label{eq:Shi17}
R_{ij}^{GS}=\sqrt{(X_j^{GS}-x_i)^2+(Y_j^{GS}-y_i)^2},\forall i\in \mathcal{N},\forall j\in \mathcal{M}.
\end{equation}

If UEs decide to offload to the GS, the data rate can be given as
\begin{equation}\label{eq:Shi18}
\begin{split}
r_{ij}=B\text{log}_2\left(1+\frac{P_{ij}^Th_{ij}^{GS}}{\sum_{k\in \mathcal{N} k\not=i}P_{kj}^Th_{kj}^{GS}+\sigma^2}\right),\\
\forall i\in \mathcal{N},\forall j\in \mathcal{M}
\end{split}
\end{equation}
where  $h_{ij}^{GS}=\alpha_j^{GS}\left(R_{ij}^{GS}\right)^{-2}$, $\alpha_j^{GS}$ is the small-scale fading component and $\left(R_{ij}^{GS}\right)^{-2}$ is the large-scale path-loss component.

Define the decision variables as $\mathbf{A}=\{a_i^L,a_{ij}^{MEC}\}$, $\forall i\in\cal N$, $\forall j\in\{\cal J,\cal K,\cal M\}$, $\mathbf{F}=\{f_i,f_{ij}\}$, $\forall i\in\cal N$, $\forall j\in\{\cal J,\cal K,\cal M\}$, the location for UAV as $\mathbf{W}=\{X_j^{UAV},Y_j^{UAV},H_j^{UAV}\}$, $ \forall j \in\cal J$, and the location for GV as $\mathbf{G}=\{X_j^{GV},Y_j^{GV}\}$, $\forall j\in\cal K$. Then, one can formulate the energy minimization optimization as
\begin{displaymath}
P1\colon \mathop{min}\limits_\mathbf{{A,F,W,G}} \mathop{\sum}\limits_{i\in \mathcal{N}} \mathop{\sum}\limits_{j\in \mathcal{\{J,K,M\}}} a_{ij}^{MEC} P_{ij}^T\frac{D_i}{r_{ij}}+ \mathop{\sum}\limits_{i\in \mathcal{N}}a_i^LP_i^E\frac{F_i}{f_i^L}
\end{displaymath}
\begin{equation}\label{eq:Shi19}
s.t.\qquad \enspace   (\ref{eq:Shi1}),(\ref{eq:Shi2}),(\ref{eq:Shi8}),(\ref{eq:Shi9}),(\ref{eq:Shi14}).
\end{equation}

One can see that Problem ($P$1) is an MINLP problem, which is hard to solve in general. This is because the admission decision $\bf{A}$ is binary while the resource allocation decision $\bf{F}$ and locations of UAV $\bf{W}$ and GV $\bf{G}$ are continuous. The major difficulty of solving  $P$1 lies in three aspects: (1) large-scale mixed integer programming optimization, (2) real-time decision-making and (3) the dynamic environment. To circumvent these hurdles, we propose a novel H2O framework by first applying FCM clustering algorithm to get the positions of GVs and UAVs. Then, DNN is trained offline by using the samples obtained from PSO with the global search. After that, DNN can be applied to make the real time decision, based on the input of fuzzy membership information, even in a fast changing environment. The important notations used throughout this paper are summarized in TABLE \ref{tab:new}.

\begin{table}[]
	\centering\makegapedcells
	\caption{Notations used throughout this paper.}
	\label{tab:new}
\begin{tabular}{|p{35pt}|p{185pt}|}
	\hline
	Notation & Description \\
	\hline
$\mathcal{N}$	& The set of UEs \\
	\hline
$\mathcal{J}$, $\mathcal{K}$, $\mathcal{M}$	&  The sets of UAVs, GVs and GSs\\
	\hline
$a_i^L$ 	& Local executing indicator \\
	\hline
$a_{i,j}^{MEC}$	& Offloading indicator \\
	\hline
 $P_{ij}^T$  	& Transmitting power \\
	\hline
$P_{i}^E$ 		&  Local execution power\\
	\hline
$D_i$  		& The transmitting data size  \\
	\hline
$F_i$	&  The required number of the CPU cycles	\\
	\hline
$f_i^L$ 	&   Local executing capacity \\
	\hline
	$r_{ij}$		&   Data rate \\
	\hline
 $h_{ij}$			& Channel quality information \\
	\hline
 $T^{req}$ 			&  The latency constraint \\
	\hline
 $f_{ij}$ 		&  Computation capacity of the $j$-th H-MEC providing to the $i$-th UE	\\
	\hline
		$\alpha_j$ & The small-scale fading component  \\
	\hline
	$R_{ij}^{-2}$	&  The large-scale path-loss component  \\
	\hline
		$\mathbf{C}$	&  Cluster center matrix \\
	\hline
	 $\tau$   	& Weighting exponent of FCM \\
	\hline
	 $c$ 	&  The number of clusters\\
	\hline
 $\varepsilon$	& Convergence threshold of FCM \\
\hline
	 $T_{FCM}$ & The maximum iteration number of FCM \\
	\hline
	  $\mathbf{U}$	& Fuzzy membership matrix  \\
	\hline
	$\boldsymbol{x}_i$ 	& Position vector of PSO \\
	\hline
	$\boldsymbol{v}_i$	&  Velocity vector of PSO\\
	\hline

$w_{max}$ 	&  The initial inertia weight of PSO \\
	\hline
$w_{min}$ 	&  The final inertia weight of PSO \\
	\hline
	
	 $\boldsymbol{p}_g$	&  The global best particle \\
	\hline
		$P$	&  The number of particles \\
	\hline
	$T_{PSO}$ & The maximum iteration number of PSO \\
	\hline

	 $\boldsymbol\theta$	& Network parameters of DNN  \\
	\hline
	$\beta$ 	&  The learning rate of DNN\\
	\hline

$\mathbf{r}_{\ell}$ 		& The $\ell$-th layer output of DNN   \\
	\hline
$\boldsymbol{u}_i$ 		&   Fuzzy membership information of the $i$-th UE\\
	\hline
 $N_c$		& The number of constraints  \\
	\hline
	$T_{CNN}$ & The maximum iteration number of CNN \\
\hline
	
\end{tabular}
\end{table}

\section{The hybrid deep learning-based online offloading framework}
\subsection{Algorithm Overview}

\begin{figure*}[htpb]
	\centering
	\includegraphics[width=18cm]{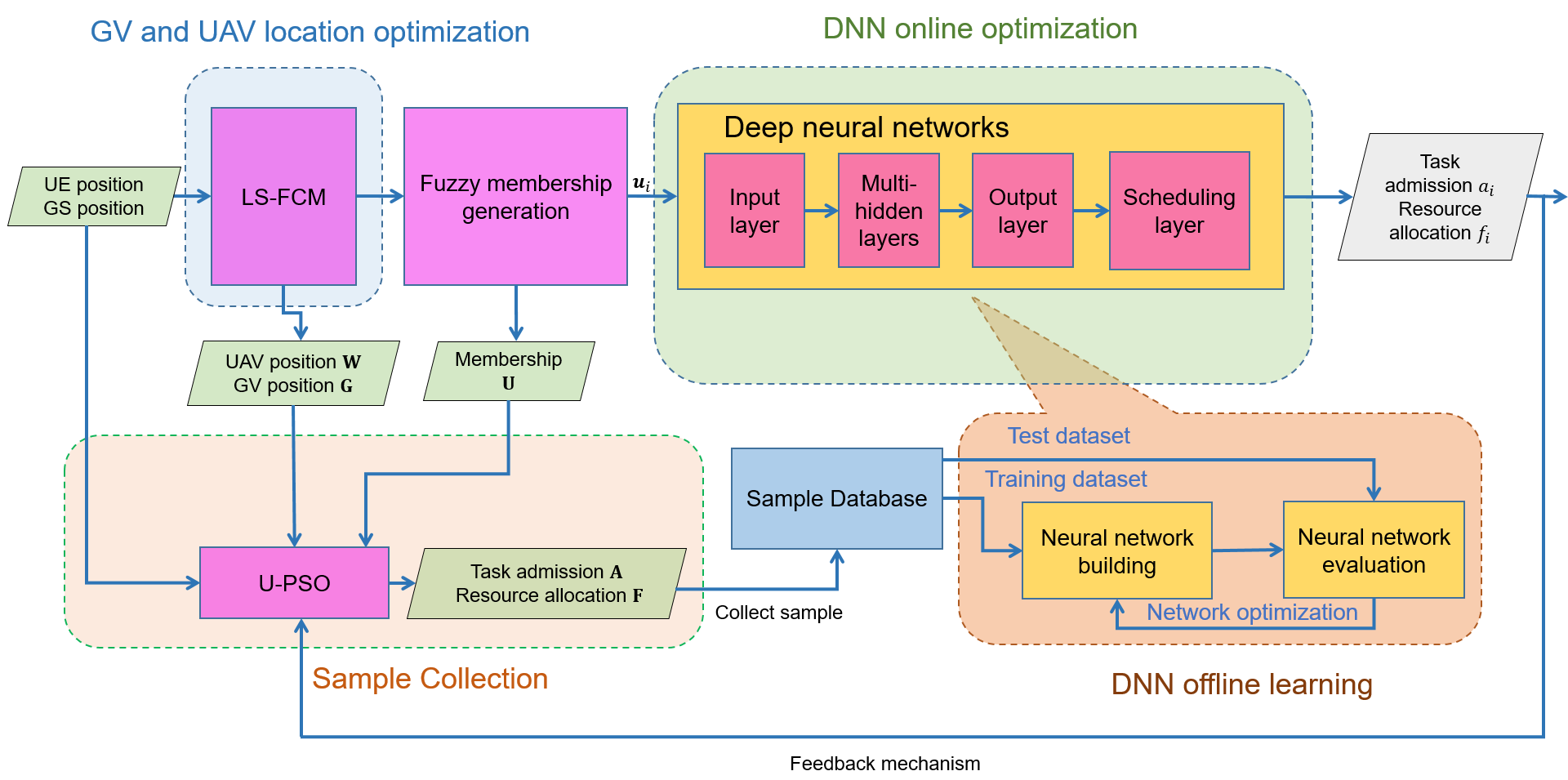}
	\caption{The proposed H2O framework.}
\label{fig:fig2}
\end{figure*}

In this section, we provide a novel algorithm named hybrid deep learning based online offloading (H2O) to solve Problem $P$1. The structure of the H2O algorithm is illustrated in Fig. \ref{fig:fig2}, which is composed of four parts: GV and UAV location optimization,
sample collection, DNN offline learning and DNN online decision. The main procedure of the H2O algorithm is divided into the offline training phase and the online optimization phase.

The offline training phase is carried out in the remote cloud server with high computational and storage capabilities. We regard the optimization problem $P$1 as the mapping function from the input system parameters to the output resource allocation solutions. We use a DDN to fit this mapping function. To this end, we need to find the resource allocation solution under certain system parameters. In other words, we need samples to train the DNN. To this end, we provide one algorithm to solve Problem $P$1 to obtain the training samples. In specific, we first adopt the LS-FCM algorithm in Subsection-B to obtain the locations of the UAVs and GVs. We then deploy the UAVs and GVs according to the calculated locations. Then, we propose a novel U-PSO algorithm in Subsection-C to obtain the offloading selection and resource allocation solution based on the novel fuzzy membership information that can capture the small-scale channel information and the relative interference among the UEs. Finally, supervised learning algorithm is used to train the DNN that can be applied for the scenarios where the number of UEs is varying.

Then, the trained DNN can be implemented for online calculation. In particular, each UE only needs to input its membership values (detailed in Subsection-C), then the DNN can output its offloading choice and computing resource allocation result. This has much lower computational complexity since it only needs to perform some simple algebraic calculations instead of solving the original optimization problem through high-complexity heuristic algorithms. In addition, during the online implementation, some results output by the DNN will be regarded as new samples and be stored in the database at the cloud.  These samples will be used for offline training and to update the DNN. This is very important since it can track the variations of the real scenarios.

In the following, we provide more details of each step of the H2O algorithm.

\subsection{Location optimization based on LS-FCM}
Clustering algorithm is always applied to determine the locations of UAVs \cite{7875131}, but it cannot be directly used in our work. First, the positions of GSs are fixed, but the positions of GVs and UAVs are varying. The conventional clustering method did not consider the case when some points are fixed. Second, the iterative process of conventional FCM only considers the distances between UEs and MECs, and does not consider CQI between them. To solve these problems, we propose a novel large-scale path-loss fuzzy c-means (LS-FCM) algorithm to locate the positions of UAVs and GVs, which has two advantages: First, it can fix some cluster centers denoted as GS positions and not allow them to participate in the iteration process. Second, it introduces the large-scale path-loss component to replace distance in clustering process.

FCM clustering is a data clustering method in which each data point belongs to a fuzzy cluster, with a degree specified by a membership grade. FCM partitions a collection of $n$ data point  $\boldsymbol{z}_i$ into fuzzy clusters. In our study, $\boldsymbol{z}_i$ is the $i$-th data point which denotes the position $(x_i,y_i)$ of the  $i$-th UE.

Specifically, the objective function of LS-FCM is expressed in the following:
\begin{equation}\label{eq:Shi20}
G=\sum_{i\in\mathcal{N}}\sum_{j\in \mathcal{\{J,K,M}\}}(\mu_{i j})^{\tau}d_{ij}^{\prime}
\end{equation}
where $d_{i j}^{\prime}=1/R_{i j}^{-2}$ denotes the large-scale path-loss component, $\mu_{i j}$ denotes the fuzzy relationship between the $i$-th UE and the $j$-th MEC. Here, $\boldsymbol{C}_j$ is denoted as the $j$-th cluster center which denotes the position of the $j$-th H-MEC (including UAVs, GVs and GSs). $\tau (\tau>1)$ is a scalar termed as the weighting exponent that controls the fuzziness of the resulting clusters.
To minimize the criterion $G$, the centroid  $\boldsymbol{C}_j$ is updated according to Eq. (\ref{eq:Shi21}), and $\mu_{ij}$ is updated according to Eq. (\ref{eq:Shi22}), respectively,
\begin{equation}\label{eq:Shi21}
\boldsymbol{C}_{j}=\frac{\sum_{i \in \mathcal{N}}\left(\mu_{i j}\right)^{\tau} z_{i}}{\sum_{i \in \mathcal{N}}\left(\mu_{i j}\right)^{\tau}}, \forall j \in\{\mathcal{J}, \mathcal{K}, \mathcal{M}\},
\end{equation}
and
\begin{equation}\label{eq:Shi22}
\mu_{ij}=\frac{1}{\sum_{k=1}^c\left(\frac{d_{ij}^{\prime}}{d_{ik}^{\prime}}\right)^{\frac{2}{\tau-1}}},
\end{equation}
where $c$ is the number of clusters.

$\mu_{ij}$ and $\boldsymbol{C}_j$ are calculated iteratively through these two equations until the FCM algorithm converges. Based on the solutions of $\mathbf{C}=[\boldsymbol{C}_j]$, we can decide the locations of UAVs and GVs. The reason why we use FCM to locate the position of UAVs and GVs is that the LS-FCM clustering algorithm aims at minimizing the total large-scale path loss of all UEs as seen in Eq. (\ref{eq:Shi20}), which implicitly reduces the system energy  shown in the objective function of Problem (19).

Finally, a hard classification is adopted by assigning each UE solely to the cluster with the highest $\mu_{ij}$. In the LS-FCM, since the GS positions are fixed, some cluster centers are set to the GS positions in advance and are not allowed to participate in the iteration process. According to the hard classification, the cluster centers are sorted in a descending order according to the total required computing resources of all UEs in this cluster. Then the centers of the cluster are assigned to GVs and UAVs based on  their available computing resources. For example, the first center of the cluster is assigned to the H-MEC with the most computing resources, the last center of the cluster is assigned to the H-MEC with the least computing resources. Please note that this is not the final computing offloading solution. This stage just provides a rough estimation of computing resource needed by all UEs in each cluster. This LS-FCM just provides the locations for each H-MEC. Although, some UEs are in one H-MEC's cluster, these UEs may not be offloaded to this H-MEC since the small-scale fading are not considered in LS-FCM. The offloading selection and computing allocation will be studied in the next subsection.

Several stopping rules can be used in LS-FCM. One is to terminate the algorithm when the maximum iteration number $T_{FCM}$ is reached or when the variation of objective function $\Delta G$ is less than threshold $\varepsilon$.

The detailed description of LS-FCM algorithm is provided in $\bf{Algorithm\enspace\ref{alg1}}$.

\begin{algorithm}
	\caption{LS-FCM algorithm}
	\label{alg1}
	\begin{algorithmic}[1]
		\REQUIRE  GS positions, UE positions, $\varepsilon$, $c$, $\tau$, $T_{FCM}$.
		\ENSURE $\bf{C}$.
		\STATE{Initialize the locations of cluster centers. Calculate $\mu_{ij}$ according to Eq. (\ref{eq:Shi22}).}
		\STATE{Fix some centers of $\bf{C}$ according to the GS positions.}
		
		\STATE{$\Delta G(1)=1,G(0)=0,t=1.$}
		\WHILE{$|\Delta G(t)|>\varepsilon$ and $t<T_{FCM}$}
		\STATE{Calculate the remaining cluster centers of matrix $\bf{C}$ using Eq. (\ref{eq:Shi21}).}
		\STATE{Calculate the objective function $G(t)$ using Eq. (\ref{eq:Shi20}).}			
		\STATE{Update each $\mu_{ij}$ using Eq. (\ref{eq:Shi22}).}
		\STATE{$\Delta G(t)=G(t)-G(t-1)$.}
		\STATE{$t=t+1$.}
		\ENDWHILE\\
	\end{algorithmic}
\end{algorithm}

\subsection{Computing offloading selection and computing resource allocation based on U-PSO}
When the locations of UAV and GV are determined by the LS-FCM algorithm, we need to optimize the computing offloading decision for each UE and the corresponding allocated computing resources based on both the large-scale path-loss and small-scale channel fading information. This topic will be studied in this subsection. Heuristic algorithms are always used to solve the complex MINLP problem and obtain high-quality global solution\cite{8490683}. However, heuristic algorithms also exhibit some issues: it takes more calculation time than traditional gradient descent method when you want to achieve similar accuracy and the convergence speed decreases considerably in the later period of evolution. So the heuristic algorithm is fundamentally infeasible for real-time system optimization under fast changing environment. In this paper, we propose a U-PSO, in which we use PSO algorithm to generate samples for DNN offline training and then use trained DNN to make online task admission and resource allocation decision.

PSO is a heuristic computation technique that was originally proposed by Eberhart and Kennedy \cite{494215}. The standard PSO can be described as follows. Consider a swarm with $P$ particles, each particle is a candidate solution and it moves in a bounded $p$-dimensional search space. The position vector and velocity vector of the $i$-th particle are represented as  $\boldsymbol{x}_i=(x_{i1},x_{i2},\cdots,x_{ip})$ and $\boldsymbol{v}_i=(v_{i1},v_{i2},\cdots,v_{ip})$, respectively. PSO explores the search space by modifying the velocity of each particle, according to two attractors: the best position found so far by the particle itself $\boldsymbol{p}_{bi}$,
and the best position identified so far by the whole swarm $\boldsymbol{p}_g$. By integrating these two attractors, the particle behavior can be modeled by using the following equations:
\begin{align}
\label{eq:Shi23}
v_{id}(t+1)=&w(t)v_{id}(t)+c_{cog}\cdot rand_1(p_{bid}(t)-x_{id}(t))\notag\\
&+c_{soc}\cdot rand_2(p_{gd}(t)-x_{id}(t))
\end{align}
\begin{equation}\label{eq:Shi24}
x_{id}(t+1)=x_{id}(t)+v_{id}(t+1)
\end{equation}
where $c_{cog}$ is the cognitive factor and  $c_{soc}$ is the social factor; $rand_1$ and $rand_2$ are generated randomly between 0 and 1; $p_{bid}(t)$ is the best position of the $i$-th particle at iteration $t$; $p_{gd}(t)$ is the best particle position among all the particles at iteration $t$; $d$=1,2,…,$p$. $w(t)$ is the inertia weight. Typically, the inertia weight is set according to \cite{Jiang2016}:
\begin{equation}\label{eq:Shi25}
w(t)=w_{max}-(w_{max}-w_{min})\cdot t/T_{pso}
\end{equation}
where $w_{max}$ is the initial inertia weight, $w_{min}$ is the final inertia weight, $T_{pso}$ is the maximum iteration number.

The energy-efficient admission of delay-sensitive tasks is a complex MINLP problem, using traditional PSO algorithm to solve this problem has three drawbacks that avoids its direct application in our considered optimization problem. Firstly, PSO algorithm often employs continuous real-valued encodings, but the admission decision matrix $\bf{A}$ is a matrix with integer elements equal to 0 or 1; Second, traditional PSO did not need to check the constraints; Third, traditional PSO initializes the particle population randomly, and it does not take advantage of the CQI information. In this regard, we propose a new U-based particle swarm optimization (U-PSO) algorithm to solve the MINLP problem efficiently.

Firstly, we improve the coding of particle. In our U-PSO algorithm, the particle can be represented as:
\begin{align}\label{eq:Shi26}
\boldsymbol{x}=&[a_1,a_2,\cdots,a_i,\cdots,a_N,f_1,f_2,\cdots,f_i,\cdots,f_N]\notag\\
={}&[\boldsymbol{x}_a,\boldsymbol{x}_f]
\end{align}
where $a_i=0$ means that the $i$-th UE decides to execute the task itself, and $a_i=k$ means that the $i$-th UE decides to offload the task to the $k$-th H-MEC, while $k\in\{1,2,\cdots,J+K+M\}$. $f_i\in\mathbb{R}$ means the allocated resource to the $i$-th UE. If $a_i=0,f_i=f_i^L$. This representation transforms the decision matrix $\bf{A}$ and resource allocation matrix $\bf{F}$ to a terse coding for PSO. Then we can round $\boldsymbol{x}_a$ part of the particle after each iteration.

Secondly, we add a constraint check step evaluating each particle. If a particle leads to constraint violation, we will assign a large penalty $\rho$ to the fitness value which is set to be equal to the objective function of (\ref{eq:Shi19}).

Thirdly, the final solution of the U-PSO algorithm mainly depends on the initial input of the algorithm. Improper selection of the initial point may cause the final solution to be stuck at a locally optimal point with low performance. Hence, it is critical to select a good initial point. In this paper, we use U-based roulette wheel selection strategy to provide high-quality initial solution for accelerating convergence. However, to implement the U-based roulette wheel, we need to know the probability of each UE to select which H-MEC to offload. One intuitive  method is to use the parameter $\mu_{ij}$ in (22) after the LS-FCM algorithm in the above subsection.  In specific, the probability of the $i$-th UE to select the $j$-th H-MEC is given by
\begin{equation}\label{eq:new2}
{p_{ij}} = \frac{{{\mu _{ij}}}}{{\sum\limits_{k\in \mathcal{\{J,K,M}\}} {{\mu _{ik}}} }}.
\end{equation}

However, this selection does not capture the instantaneous small-scale channel information and interference among the UEs. To resolve this issue, we provide a novel fuzzy membership matrix based method. It applies the small-scale fading and interference information to generate dynamic membership information, so it can output a novel fuzzy membership matrix, which includes the CQI and interference information for UEs. The fuzzy membership matrix is very important in our framework. It can not only be used here for generating the initial point of the U-PSO algorithm, but also serves as the input as the DNN discussed in the next subsection.

we define the novel fuzzy membership matrix $\mathbf{U}=[u_{ij}]$ including the small-scale fading and interference information, then we update the fuzzy membership matrix in the dynamically changing environment. The membership degree  $u_{ij}$ is expressed as following:
\begin{equation}\label{eq:new}
u_{ij}=\frac{1}{\sum_{k=1}^c\left(\frac{h_{ij}^{\prime}}{h_{ik}^{\prime}}\right)^{\frac{2}{\tau-1}}},
\end{equation}
where $h_{ij}^{\prime}=\left(\gamma \sum_{k \in \mathcal{N}, k \neq i} P_{kj}^T\alpha_jR_{k j}^{-2}\right) / P_{ij}^T\alpha_jR_{i j}^{-2}$ is an enhanced version of  $d_{i j}^{\prime}$, which includes the small-scale fading component  $\alpha_j$ for the $j$-th MEC and the referenced interference information $\sum_{k \in \mathcal{N}, k \neq i} P_{kj}^T\alpha_jR_{k j}^{-2}$, $\forall j\in \{\mathcal{K},\mathcal{M}\}$, and $\gamma$ is the trade-off factor. The fuzzy membership matrix  $\mathbf{U}$ can reflect the changes of channel and the interferences of environment in real time.

By using the updated dynamic membership information defined in Eq. (\ref{eq:new}), we now can obtain the probability of UE $i$ to select H-MEC $j$ is given by
\begin{equation}\label{eq:new3}
{p_{ij}} = \frac{{{u _{ij}}}}{{\sum\limits_{k\in \mathcal{\{J,K,M}\}} {{u _{ik}}} }}.
\end{equation}

\begin{figure}[htpb]
	\centering
	\includegraphics[width=8cm]{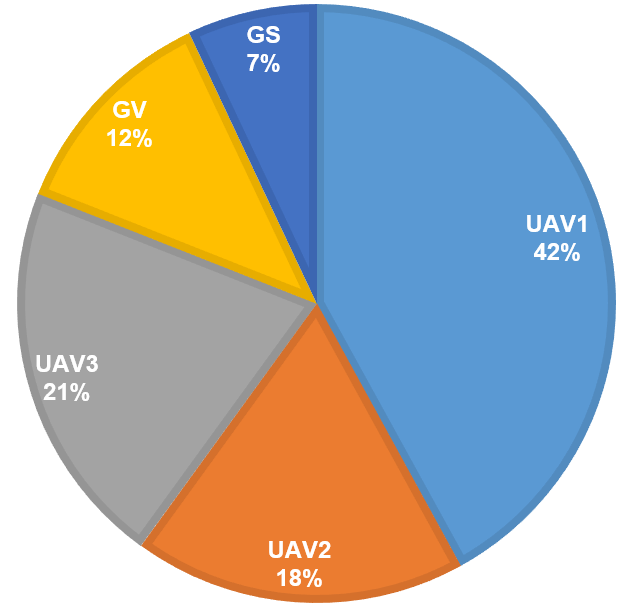}
	\caption{Roulette wheel selection strategy.}
	\label{fig:fig3}
\end{figure}

With the probability values, we can now employ the U-based roulette wheel to initialize the initial solution. For example, suppose there are five H-MECs that UE $i$ can choose to offload (three UAVs, one GV and one GS), in Fig. \ref{fig:fig3} the circumference of the roulette wheel is equal to one. UAV1 is the most fit individual and has the largest probability to be selected, whereas GS and GV are the least fit and have smaller intervals within the roulette wheel. To select a H-MEC, a random number is generated in the interval [0,1], and the H-MEC whose segment spans the random number is selected and UE $i$ will offload its task to this H-MEC. This process is repeated until all UEs have selected their H-MECs. Then based on the offloading decision, the resource of each H-MEC is allocated evenly to its associated UEs. The process of U-PSO algorithm is present in $\bf{Algorithm\enspace\ref{alg2}}$.

The fuzzy membership matrix $\mathbf{U}$ which captures small-scale fading and mutual interference information plays a key role in our H2O framework. The fuzzy membership matrix $\mathbf{U}$ can not only guide the initialization of U-PSO, but also provide a concise representation of the relationship between UEs and H-MECs, which will be a perfect input of DNNs.
\begin{algorithm}
	\caption{U-PSO algorithm}
	\label{alg2}
	\begin{algorithmic}[1]
		\REQUIRE  H-MEC positions, UE positions, $P$, $\rho$, $w_{max}$, $w_{min}$, $T_{pso}$.
		\ENSURE $\boldsymbol{p}_g$.
		\STATE{Calculate the dynamic fuzzy membership matrix $\bf{U}$ using Eq. (\ref{eq:new}).}
		\STATE{Initialize the particle population by using the U-based roulette wheel selection strategy according to $\bf{U}$.}
		\WHILE{$t<T_{pso}$}
		\STATE{Calculate the fitness value of each particle using Eq. (\ref{eq:Shi19}).}
		\STATE{Check the constraints and assign the penalty $\rho$ to the particle with constraint violations.}
		\STATE{Save the global best particle $\boldsymbol{p}_g$ of the swarm.}			
		\STATE{Save the individual best $\boldsymbol{p}_{bi}$ of each particle.}
		\STATE{Update the velocity of each particle using Eq. (\ref{eq:Shi23}) and Eq.  (\ref{eq:Shi25}).}
		\STATE{Update the position of each particle using Eq. (\ref{eq:Shi24}).}
		\STATE{Round the $\boldsymbol{x}_a$ part of each particle.}
		\STATE{$t=t+1$.}
		\ENDWHILE
	\end{algorithmic}
\end{algorithm}

\subsection{Offline DNN training and online implementation}
Given the locations of UAVs and GVs, the Problem  $P$1 can be regarded as an unknown function mapping from the fuzzy membership matrix $\bf{U}$ to the optimal task admission $\mathbf{A}=[a_1,a_2,\cdots,a_i,\cdots,a_N]$ and resource allocation $\mathbf{F}=[f_1,f_2,\cdots,f_i,\cdots,f_N]$, namely.
\begin{equation}\label{eq:Shi27}
\mathcal{F}\colon\mathbf{U}\in\mathbb{R}^{{N}\times{(J+K+M)}}\to[\mathbf{A},\mathbf{F}]\in \mathbb{R}^{N\times2}
\end{equation}
where $a_i=0$ means the $i$-th UE decides to execute the task itself, $a_i=k$ means the $i$-th UE decides to offload the task to the $k$-th H-MEC, while $k\in\{1,2,\cdots,J+K+M\}$. $f_i\in\mathbb{R}$ means the allocated resource to the $i$-th UE. If $a_i=0,f_i=f_i^L$.

Considering that DNN is a universal function approximator, we use DNN to learn the unknown function mapping. The DNN with $L$ layers describes a mapping $f(\mathbf{r}_0;\boldsymbol{\theta})\in \mathbb{R}^{N_0}\to\mathbf{r}_L\in\mathbb{R}^{N_L}$ of an input vector $\mathbf{r}_0\in\mathbb{R}^{N_0}$ to an output vector $\mathbf{r}_L\in\mathbb{R}^{N_L}$ through $L$ iterative processing steps:
\begin{equation}\label{eq:Shi28}
\mathbf{r}_{\ell}=f_{\ell}(\mathbf{r}_{\ell-1};{\boldsymbol\theta_{\ell}}),\ell=1,\cdots,L
\end{equation}
where $\mathbf{r}_{\ell}\in\mathbb{R}^{N_{\ell}}$ is the output of the $\ell$-th layer.  $\boldsymbol\theta_{\ell}$ is a set of parameters of the $\ell$-th layer. The $\ell$-th layer is called fully-connected if $f_{\ell}(\bf{r}_{\ell-1};\boldsymbol\theta_\ell)$ has the form
\begin{equation}\label{eq:Shi29}
f_{\ell}(\mathbf{r}_{\ell-1};{\boldsymbol\theta_{\ell}})=\sigma(\mathbf{W}_{\ell}  \mathbf{r}_{\ell-1}+\mathbf{b}_{\ell})
\end{equation}
where $\mathbf{W}_{\ell}\in\mathbb{R}^{N_{\ell}\times N_{\ell-1}}$ is the weights of the $\ell$-th layer, $\mathbf{b}_{\ell}\in\mathbb{R}^{N_{\ell}}$ is the thresholds of the $\ell$-th layer. The set of parameters for this layer is $\boldsymbol\theta_{\ell}=\{\mathbf{W}_{\ell},\mathbf{b}_{\ell}\}$.  We use $\boldsymbol\theta=\{\boldsymbol\theta_1,\cdots,\boldsymbol\theta_L \}$ to denote the set of all parameters of the DNN. $\sigma(\cdot)$ is an activation function. Some commonly used activation functions are ReLU, tanh, sigmoid and softmax listed in TABLE \ref{tab:table1}\cite{Schmidhuber2015Deep}.
\begin{table}[]
		\centering\makegapedcells
	\caption{activation functions of DNN.}
	\label{tab:table1}
	\begin{tabular}{|l|l|l|}
		\hline
		Name   & $\sigma(v_i)$  & Range \\ \hline
		ReLU   & $max(0,v_i)$  & $[0,\infty)$ \\ \hline
		tanh   & $\frac{e^{v_i}-e^{-v_i}}{e^{v_i}+e^{-v_i}}$  & $(-1,1)$ \\ \hline
		sigmoid   & $\frac{1}{1+e^{-v_i}}$  &  $(0,1)$ \\ \hline
	    softmax & $\frac{e^{v_i}}{\sum_je^{-v_i}}$ & $(0,1)$ \\ \hline
	\end{tabular}
\end{table}

In order for the DNN to learn the desired input-output mapping, it is necessary to tune the parameters $\boldsymbol\theta$ in a supervised learning fashion. We define the loss function $\mathcal L(\boldsymbol{p}^{(n)},\boldsymbol{r}_L^{(n)})$ that is used to measure the loss between the desired output ${\boldsymbol{p}^{(n)}}$ and the actual output $\boldsymbol{r}_L^{(n)}$. The goal of the training process is to tune the network parameters $\boldsymbol\theta$ in order to minimize the average loss, defined by
\begin{equation}\label{eq:Shi30}
L(\boldsymbol{\theta})=\frac{1}{N_t}\sum_{n=1}^{N_t}\mathcal{L}(\boldsymbol{p}^{(n)},\boldsymbol{r}_{L}^{(n)})
\end{equation}
where $N_t$ is the number of DNN samples. Several relevant loss functions are represented in TABLE \ref{tab:table2} \cite{8054694}.
\begin{table}[]
	\centering\makegapedcells
	\caption{ loss functions of DNN}
	\label{tab:table2}
	\begin{tabular}{|l|l|}
		\hline
	 Name  & $\mathcal{L}(\boldsymbol{p},\boldsymbol{r})$ \\ \hline
  	MSE  & $||\boldsymbol{p}-\boldsymbol{r}||_2^2$ \\ \hline
	Categorical cross-entropy  & $-\sum_jp_j\log(r_j)$ \\ \hline
	\end{tabular}
\end{table}	
This minimization problem can be tackled by (stochastic) gradient descent methods, i.e. iteratively updating the parameters $\boldsymbol\theta=\{\boldsymbol{W},\boldsymbol{b}\}$ according to the formulas:
\begin{equation}\label{eq:Shi31}
\boldsymbol{\theta}(t+1)=\boldsymbol{\theta}(t)-\beta\nabla L(\boldsymbol{\theta}(t))
\end{equation}
where $\beta$ is the learning rate, and the gradients are conveniently estimated based on random subsets of the complete training set, called mini-batches and leveraging the back-propagation algorithm \cite{Schmidhuber2015Deep}.

Once the parameters $\boldsymbol{W}$ and $\boldsymbol{b}$ to be used are determined as a result of the training process, the DNN is configured and able to compute task admission and resource allocation directly. This means that once the small-scale channel information is changing, the task admission and resource allocation of UEs are updated by a simple feedforward computing of the DNN through performing some simple algebraic calculations, instead of solving Problem $P$1 through high-complexity heuristic algorithms. This yields a significant complexity reduction.

However, there are still two open problems in the design of DNN model for our system. First, the number of UEs accessing the network is time-varying, so the size of $\bf{U}$ is ever-changing and we cannot obtain a group of unified samples with the same dimensionality. Second, the Problem $P$1 is a constrained optimization problem, but traditional DNN cannot guarantee the task constraints. With this regard, we propose a novel DNN with a scheduling layer. The main improvements of this new network are listed as follows.

Firstly, to model an efficient DNN for large-scale UEs, we change function mapping $\mathcal F$ to $\mathcal F1$:
\begin{equation}\label{eq:Shi33}
\mathcal{F}1\colon \boldsymbol{u}_i\in\mathbb{R}^{(J+K+M)}\to[a_i,f_i]\in\mathbb{R}^2,\forall i\in \mathcal{N}
\end{equation}
where $\boldsymbol{u}_i=[u_{i1}, u_{i2},...,u_{ic}]$ represents the membership vector of the $i$-th UE, $a_i$ and $f_i$ represents the task admission value and resource allocation value of the $i$-th UE, respectively. $a_i=0$ means the $i$-th UE decides to execute the task itself, $a_i=k$ means the $i$-th UE decides to offload the task to the $k$-th H-MEC, while $k\in\{1,2,\cdots,J+K+M\}$, $f_i\in\mathbb{R}$ means the allocated resource to the $i$-th UE. If $a_i=0, f_i=f_i^L$. This modification has the following benefits. The input dimension of the DNN only depends on the number of H-MECs and is not related to the number of UEs. In general, the number of access points or H-MECs changes much slower than the number of UEs accessing the network. Hence, our DNN can be applied in a long time once it has been trained, which is more practical. Reducing the dimension of input data can also reduce the training complexity, which is suitable for large-scale networks with large number of UEs.

Secondly, in order to guarantee task constraints, we propose a novel DNN structure with a scheduling layer after DNN training. As shown in Fig. \ref{fig:fig4}, this scheduling layer includes a constraint layer and a decision layer, the constraint layer is used to check whether the output $a_i$ and $f_i$ satisfy the constraints or not and each node in this layer represents a constraint check. The output of the $j$-th node in the constraint layer can be represented as:
\begin{equation}\label{eq:Shi34}
\mathbf{r}_{L+1,j}=g_{j}(a_i,f_i)
\end{equation}
where $g_{j}$ is the $j$-th constraint check function. If the output layer of the DNN satisfy the constraint, the function outputs ‘1’ to the next layer, else outputs ‘0’ to the next layer.

The node in the decision layer is labeled as $\Pi$, indicating that they play the role of a simple multiplier. If the output layer of the DNN doesn’t satisfy all constraints, the final output $\mathbf{r}_{L+2}=0$, which means UE execute the task locally ($a_i=0, f_i=f_i^L$), else the final output of the DNN is $\mathbf{r}_{L+2}=\mathbf{r}_{L}$. The decision layer can be represented as:
\begin{figure*}[htpb]
	\centering
	\includegraphics[width=18cm]{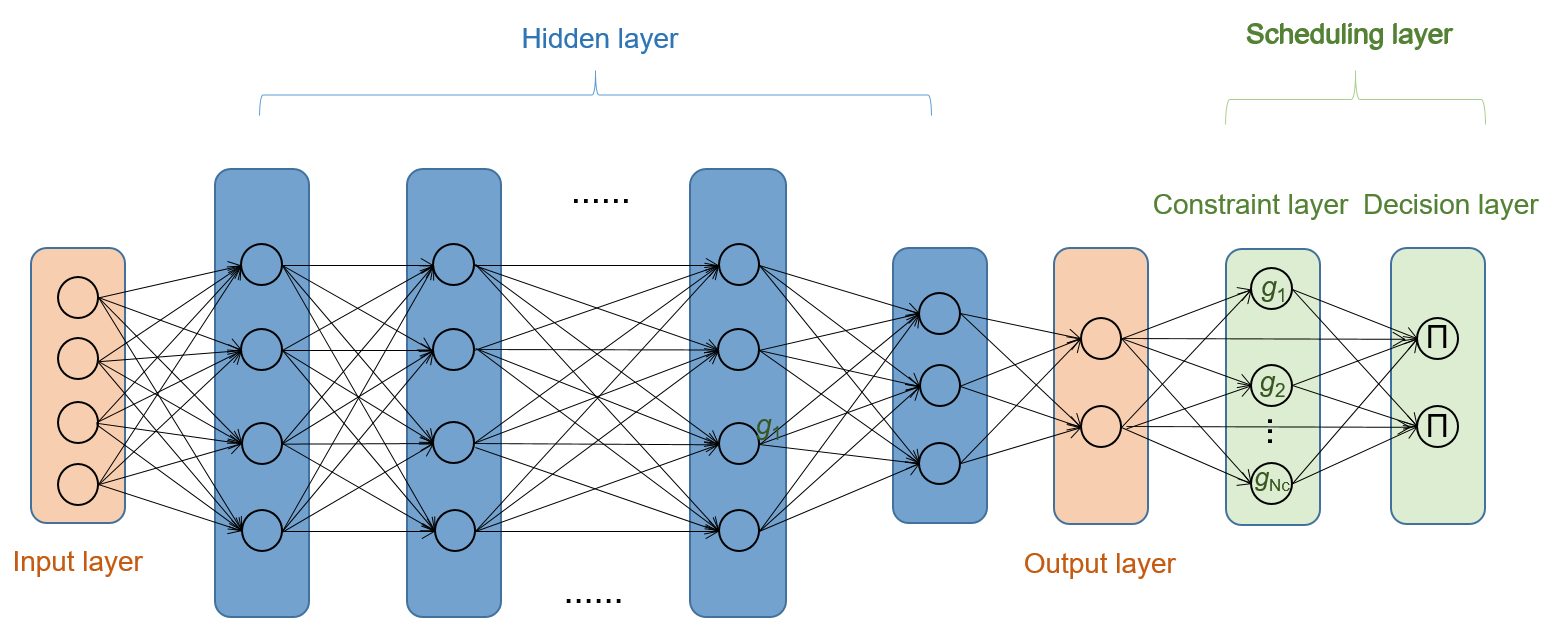}
	\caption{DNN with the scheduling layer.}
	\label{fig:fig4}
\end{figure*}
\begin{equation}\label{eq:Shi35}
\mathbf{r}_{L+2}=\mathbf{r}_{L}\cdot\Pi_{j=1}^{N_c}\mathbf{r}_{L+1,j}
\end{equation}
where $N_c$ is the number of constraints. The detailed process of the DNN with the scheduling layer is shown in $\bf{Algorithm\enspace\ref{alg3}}$.

The online implementation of the algorithm is as follows. Once the DNN is trained by the supervised learning method. Each UE only needs to input its fuzzy membership values into the DNN, and it will output the resource allocation solution with some simple algebraic calculations, which is much faster than the high-complexity heuristic algorithms. Once one new UE is accessing the network, we need to execute the FCM algorithm to obtain the new locations of the H-MECs and then this UE calculates its fuzzy membership values based on the small-scale channel information and instantaneous interference power. Then, this UE inputs the calculated membership values into the trained DNN, and it will output the resource allocation solution. Hence, our proposed algorithm is very suitable for dynamic scenarios where the number of UEs is varying, which is more practical for implementation.

In addition, the feedback mechanism is applied to H2O framework for online implementation. In practical terms, we use a differential check method to realize the feedback mechanism. When the new data is inputted to the DNN, we check the difference between the new data and the existing samples in the database, which is evaluated by the Euclidean distance. If there is a big difference between the new data and the existing samples, the new data will be fed back to the U-PSO and resolved as a new sample.
\begin{algorithm}
  \caption{DNN with a scheduling layer algorithm}
  \label{alg3}
  \begin{algorithmic}[1]
  \REQUIRE $\bf{U}$, $\beta$, $N$, $T_{CNN}$.
  \ENSURE $\boldsymbol\theta$, $\mathbf{r}_{L+2}$.
  \\
  $\bf{Training\; stage:}$\\
  \STATE{Randomly initialize network parameters $\boldsymbol\theta$.}
  \WHILE{$t<T_{CNN}$}
  \STATE{Calculate the feedforward of DNN according to Eqs. (\ref{eq:Shi28})-(\ref{eq:Shi29}) for all layers.}
  \STATE{Calculate the loss function according to Eq. (\ref{eq:Shi30}).}
  \STATE{Update of DNN according to Eq. (\ref{eq:Shi31}).}		
  \STATE{$t=t+1$}.
  \ENDWHILE
  \STATE{Serialize $\bf{U}$ into a set of $\boldsymbol{u}_i$ according to the UE number $N$.}\\
  $\bf{Decision\; stage:}$\\
  \WHILE{$i<N$}
  \STATE{Calculate the output $a_i$ and $f_i$ of the DNN based on the trained $\boldsymbol\theta$ according to the input $\boldsymbol{u}_i$. }
  \STATE{Check the output of DNN by the constraint layer according to Eq. (\ref{eq:Shi34}).}
  \STATE{Calculate the decision output  $\mathbf{r}_{L+2}$ by the decision layer according to Eq. (\ref{eq:Shi35}).}
  \STATE{$i=i+1$}.
	\ENDWHILE
  \end{algorithmic}
\end{algorithm}

\section{Simulation results}
In this section, we use simulations to evaluate the performance of the proposed H2O algorithm. In all simulations, we use three UAVs, one GV and one GS. Other parameters used in the simulations are summarized in TABLE \ref{tab:table3}, unless otherwise specified. The parameters of the LS-FCM method are chosen as follows: $\tau=2$, $\varepsilon=0.0001$, $T_{FCM}=100$; The parameters of the U-PSO method are chosen as follows: $P=10, \rho=100, w_{max}=0.9, w_{min}=0.4, T_{PSO}=100$; The parameters of the DNN method are chosen as follows: $\beta=0.4,  T_{DNN}=500$, the activation function is set to ReLU and function is set to MSE.

Apart from the proposed task admission algorithm, we also simulate the following algorithms for comparison purposes.

\begin{itemize}
\item Random offloading (Random): the task admission is decided randomly for each UE. If the computational resources of the allocated H-MEC are insufficient, UE execute the task locally.
\item Greedy offloading (Greedy): all UEs offload the task to the nearest H-MEC, if the computational resources of the allocated H-MECs are insufficient, the UEs who need more computational resources of the H-MECs execute the task locally.
\item Local execution (Local): There is no offloading. All tasks are executed locally.
\item PSO offloading (PSO): the task admission is optimized by the U-PSO method.
\end{itemize}

\begin{figure}[htpb]
	\centering
	\includegraphics[width=9cm]{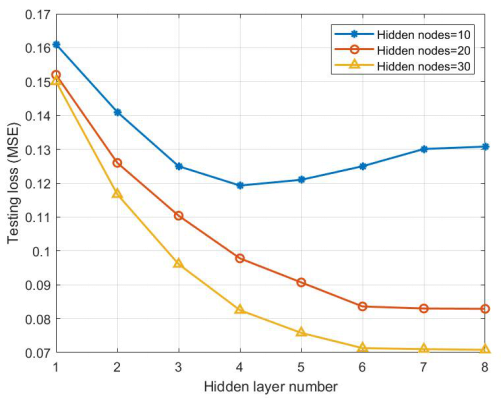}
	\caption{The comparison of testing loss of DNN with different hidden layer number and different hidden nodes number.}
	\label{fig:fig6}
\end{figure}

\begin{table}[]
	\centering\makegapedcells
	\caption{Simulation parameters}
	\label{tab:table3}
	\begin{tabular}{|l|l|}
		\hline
		$\bf{Parameters}$  & $\bf{Assumptions}$ \\ \hline
		Macrocell Radius   & $100$m \\ \hline
		GS Coordinate  & [50,50] \\ \hline
		Bandwidth $B$  & 1MHz \\ \hline
		Transmitting Power $P_{ij}^T$  & 1W \\ \hline
		Input Data Size  ${D_i}$  & 100kB \\ \hline
		Required Number of CPU Cycles $f_i^{L}$  & $10^9$ cycles/s \\ \hline
		Local Computational Capability $F_{max}^{L}$  & $10^9$ cycles/s \\ \hline
		Latency Request $T^{req}$  & 2s \\ \hline
		UAV Altitude ${H^{UAV}}$  & 20m \\ \hline
		Remote Computational Capability (UAV)  & $10^{10}$ cycles/s \\ \hline
		Remote Computational Capability (GV)  & $10^{11}$ cycles/s \\ \hline
		Remote Computational Capability (GS)  & $10^{12}$ cycles/s \\ \hline
	\end{tabular}
\end{table}	
All sample vectors in the dataset are split randomly into a training set and a testing set. 80 percent of the sample vectors are assigned into the training set, while 20 percent of the sample vectors are assigned into the testing set. Then cross validation method \cite{jiang2018electrical} is used to evaluate the performance of DNN.
In Fig. \ref{fig:fig6}, we compare the performance of DNN with different number of hidden nodes in hidden layer, as the number of hidden layers varies from 1 to 8. With the increase of hidden layer number, the testing loss of DNN with 20 hidden nodes and 30 hidden nodes first decrease and then stabilize when the hidden layer number is above 6. This is because increasing hidden nodes and hidden layers can enhance the learning ability of DNN and allow the DNN to learn more information. The DNN with 30 hidden nodes of each hidden layer achieves the minimum testing loss when the hidden layer number is above 6. Notice that the testing loss of DNN with 10 hidden nodes first decrease and then increase when the hidden layer number is above 4. This is due to the concept of overfitting and the DNN cannot improve its learning ability when the hidden nodes is too less. For this reason, in the follow simulations, we set the hidden layer number equal 6, the hidden nodes number equal 30.

\begin{figure}[htpb]
	\centering
	\includegraphics[width=9cm]{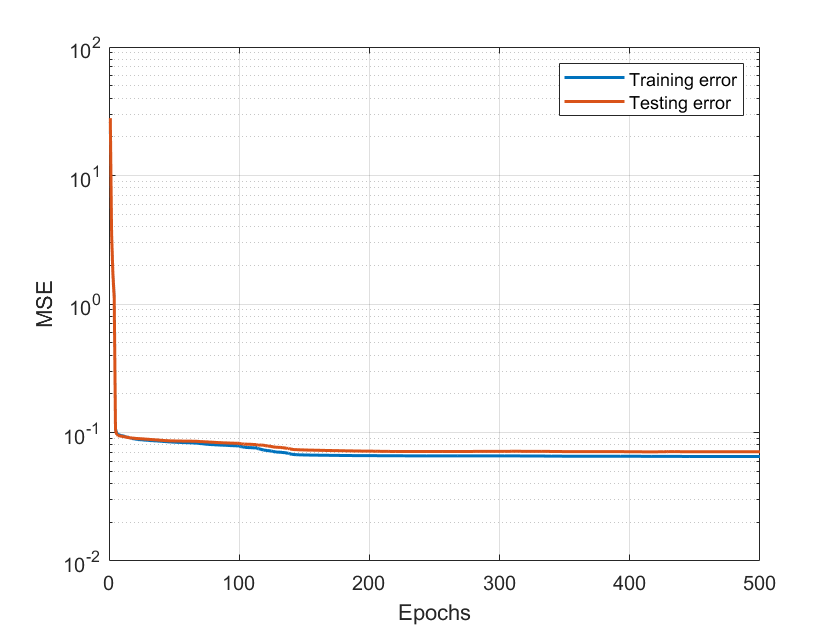}
	\caption{The testing loss and training loss of DNN.}
	\label{fig:fig7}
\end{figure}

\begin{figure}[htpb]
	\centering
	\includegraphics[width=9cm]{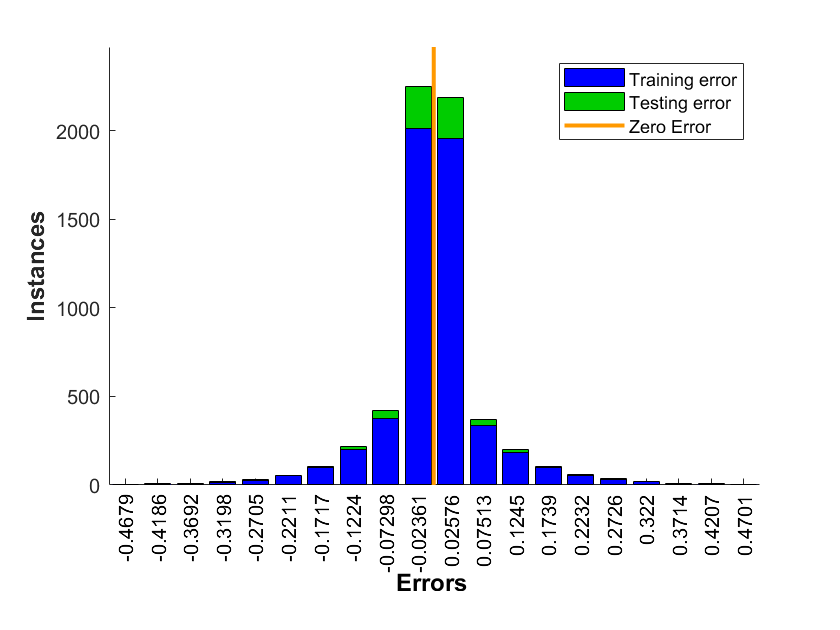}
	\caption{The training and testing error distribution of all samples for the DNN.}
	\label{fig:fig8}
\end{figure}

In Fig. \ref{fig:fig7}, we plot the training loss and testing loss of DNN. It is clearly seen that the testing loss declines sharply at the beginning of procedure, then decreases slowly and stabilize at around 0.07 when $t$ is above 300. Meanwhile, the training loss gradually decreases and stabilizes at around 0.06, whose value is less than the testing loss curve. In Fig.  \ref{fig:fig8}, we further study the error distribution of all samples for the DNN. We see that 70\% of the training errors are less than 0.025 and the maximum error is less than 0.47. The same situation applies to the error distribution of the testing samples. The simulation results demonstrate the effectiveness of the proposed DNN which can quickly converge to an optimal decision model for our offloading problem.

\begin{figure}[htpb]
	\centering
	\includegraphics[width=9cm]{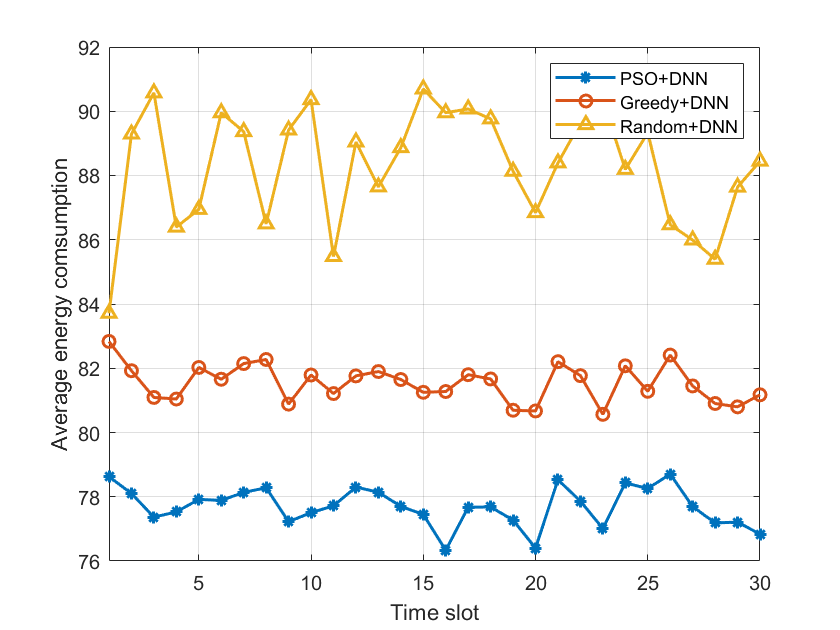}
	\caption{The comparison of average energy consumption of DNN trained by different samples under varying time slots.}
	\label{fig:fig9}
\end{figure}

In Fig. \ref{fig:fig9}, we compare the performance of DNN trained by different samples under varying time slots. Before the evaluation, DNN has been trained with 10000 independent samples generated from different algorithms (the proposed PSO, Greedy and Random), and the error curve has converged. We see that DNN achieves optimal performance with the samples generated from the proposed PSO (PSO+DNN), and significantly outperforms the DNN with the greedy offloading samples (Greedy+DNN) and the random offloading samples (Random+DNN). This is because the performance of DNN depends on the quality of samples. The proposed PSO can solve the offloading problem by a heuristic global search and achieve a high quality global solution, which guarantee the good performance of DNN from the learning stage.

We then evaluate the performance of 30 independent random scenarios for time slots prediction. We see that DNN with the samples generated from the proposed PSO can achieve the best performance under all time slots. This is because the trained DNN has good generalization performance, and can make the offloading decision continuously in fast changing environments.

\begin{figure}[htpb]
	\centering
	\includegraphics[width=9cm]{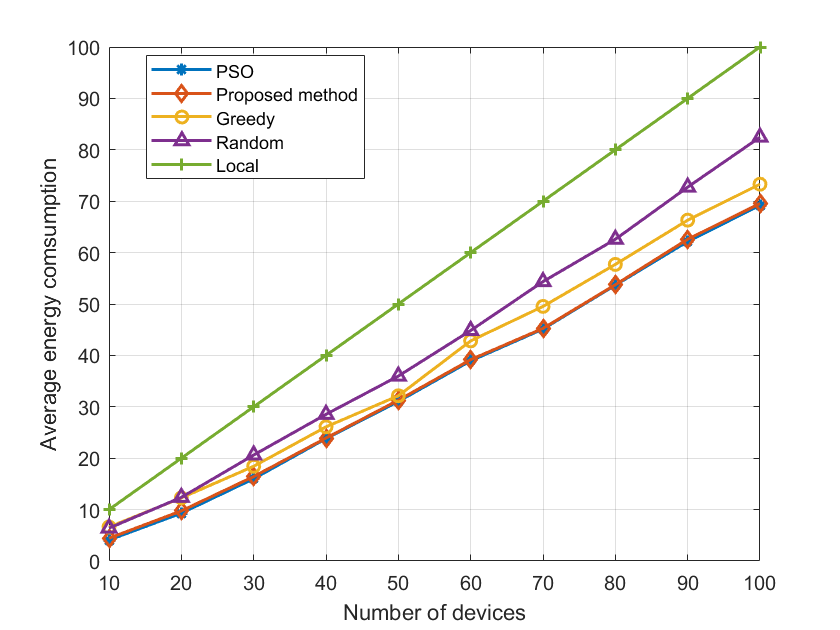}
	\caption{The comparison of average energy consumption as the number of devices varies from 10 to 100.}
	\label{fig:fig10}
\end{figure}

In Fig. \ref{fig:fig10}, we compare the average energy consumption between the proposed method, PSO, Greedy, Random and Local, as the number of devices varies from 10 to 100. As shown in the Fig. \ref{fig:fig10}, the average energy consumptions of all methods increase gradually while the number of devices increases. PSO method achieves the lowest average energy consumption. The proposed method provides almost the same energy consumption compared as PSO. This is because the proposed method is a DNN model trained by a set of high quality global solutions generated by PSO and construct a nonlinear mapping from UE and MEC information to offloading decision. Greedy method saves more energy than Random method. The energy consumption of Local is highest, since Local does not admit any devices.

\begin{figure}[htpb]
	\centering
	\includegraphics[width=9cm]{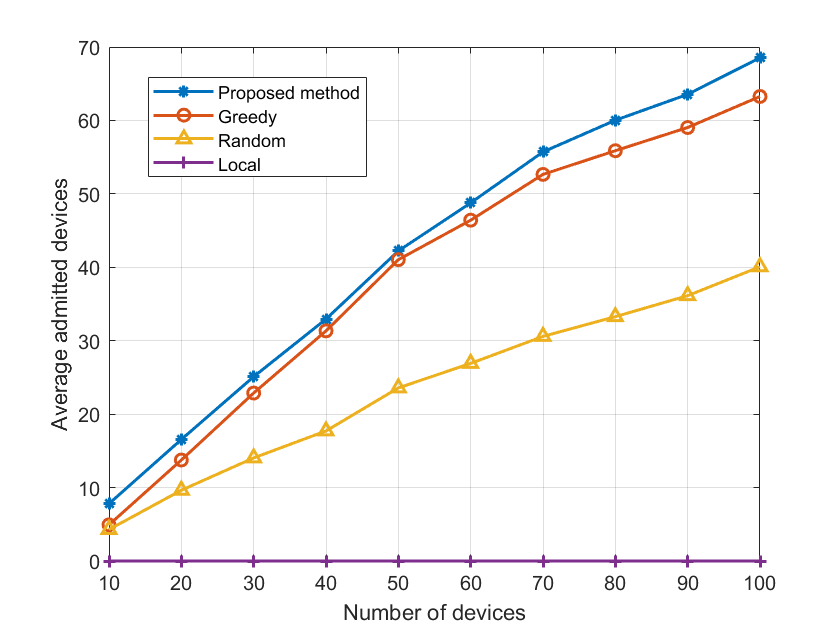}
	\caption{The comparison of average admitted devices as the number of devices varies from 10 to 100.}
	\label{fig:fig11}
\end{figure}

\begin{figure}[htpb]
	\centering
	\includegraphics[width=9cm]{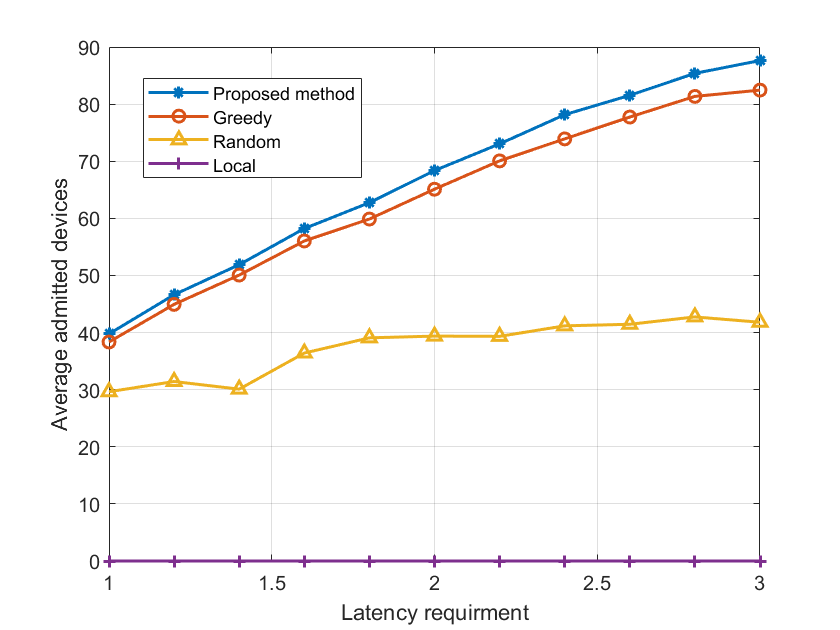}
	\caption{The comparison of average admitted devices as $T^{req}$ varies from 1s to 3s.}
	\label{fig:fig12}
\end{figure}

Fig. \ref{fig:fig11} shows the number of admitted devices (the devices who decide to offload the tasks) achieved by the proposed method, Greedy, Random and Local, where the number of devices range from 10 to 100. We can see that the proposed method admit the most devices for offloading. Greedy can admit more devices for offloading than Random. In contrast, Local does not admit any device. Fig. \ref{fig:fig12} shows the number of admitted devices achieved by the proposed method, Greedy, Random and Local, where $T^{req}$ ranges from 1s to 3s. We can see the similar results again, there may be some interesting conclusions that, for achieving the goal of minimum energy, the number of admitted devices increases as the number of devices varies from 10 to 100 or $T^{req}$ varies from 1s to 3s.

\begin{figure}[htpb]
	\centering
	\includegraphics[width=9cm]{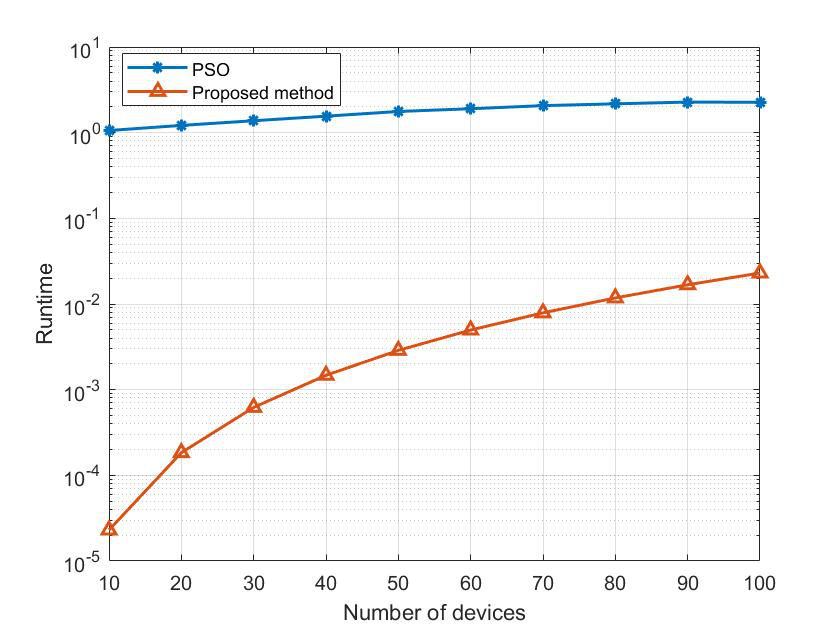}
	\caption{The comparison of runtime between PSO and the proposed method as the number of devices varies from 10 to 100.}
	\label{fig:fig13}
\end{figure}

Finally, we evaluate the computation time of the proposed algorithm. The computational complexity of all algorithms greatly depends on the number of uses. Fig. \ref{fig:fig13} compares runtime between the proposed method and PSO. We can see the proposed method is superior in terms of time efficiency. In particular, it generates an offloading action in less than 0.03 second when $N$ equals to 100, while PSO takes 80 times longer CPU time. Overall, trained DNN achieves similar performance as PSO algorithm but requires substantially less CPU time. This makes real-time offloading and resource allocation viable for H-MEC networks in fast changing environment.
\section{Conclusion}
In this paper, we have proposed a hybrid deep learning based online offloading algorithm, H2O, to minimize the sum energy consumption of UEs in a H-MEC network with binary computation offloading. The framework includes three artificial intelligence algorithms: an LS-FCM method is used to locate the GVs and UAVs, a U-PSO is used to solve the MINLP problem and provide high quality samples to DNN, a DNN with a scheduling layer is applied to make the task admission and resource allocation decision in real time. Compared to conventional optimization methods, the proposed H2O algorithm completely removes the need of solving MINLP problems in the decision period of DNN. Simulation results show that H2O achieves similar near-optimal performance as heuristic methods but reduces the CPU time by more than several orders of magnitude, making real-time system optimization feasible for the H-MEC networks in fast-changing environment.


%

%




\bibliographystyle{ieeetran}
\bibliography{bare_jrnl_bobo}

\end{document}